\documentclass[10pt,conference]{IEEEtran}
\usepackage{cite}
\usepackage{amsmath,amssymb,amsfonts}
\usepackage{mathptmx}
\usepackage{mathtools}
\usepackage{amsmath}
\usepackage{color}
\usepackage{xspace}
\usepackage{listings}
\usepackage{comment}
\usepackage{subcaption}
\usepackage{multirow}
\usepackage{color}
\usepackage{xcolor,colortbl}
\usepackage{algorithmic}
\usepackage{graphicx}
\usepackage{textcomp}
\usepackage{xcolor}
\usepackage[hyphens]{url}
\usepackage{fancyhdr}
\usepackage{hyperref}
\usepackage{dblfloatfix}
\usepackage[normalem]{ulem}
\usepackage[hang,flushmargin]{footmisc}

\let\svthefootnote\thefootnote
\newcommand\freefootnote[1]{%
  \let\thefootnote\relax%
  \footnotetext{#1}%
  \let\thefootnote\svthefootnote%
}

\def\BibTeX{{\rm B\kern-.05em{\sc i\kern-.025em b}\kern-.08em
    T\kern-.1667em\lower.7ex\hbox{E}\kern-.125emX}}

\newcommand{\rev}{}

\newcommand{\hpcayear}{2024}

\newcommand{\sysname}{PIMSAB\xspace} 
\newcommand{\gpuspeedup}{3$\times$\xspace}
\newcommand{\gpuenergy}{4.2$\times$\xspace}
\newcommand{\dcspeedup}{3.7$\times$\xspace}
\newcommand{\simdramspeedup}{3.88$\times$\xspace}

\newcommand{\jian}[1]{\textcolor{blue}{\small [Jian: #1]}}

\renewcommand{\paragraph}[1]{\noindent \textbf{#1:}}

\newcommand{\hpcasubmissionnumber}{561}

\title{\sysname: A \underline{P}rocessing-\underline{I}n-\underline{M}emory \underline{}System with \underline{S}patially-\underline{A}ware Communication and \underline{B}it-Serial-Aware Computation}

\def\hpcacameraready{} 

\newcommand\hpcaauthors{
Aman Arora$^*\dagger\Theta$, 
Jian Weng$^*\ddagger\Delta$, 
Siyuan Ma$\dagger$, 
Tony Nowatzki$\ddagger$, 
Lizy K. John$\dagger$}
\newcommand\hpcaaffiliation{
University of Texas Austin$\dagger$, 
University of California Los Angeles$\ddagger$,\\
Arizona State University$\Theta$,
King Abdullah University of Science and Technology$\Delta$
}
\newcommand\hpcaemail{aman.kbm@utexas.edu, jian.weng@ucla.edu}

\author{
  \ifdefined\hpcacameraready
    \IEEEauthorblockN{\hpcaauthors{}}
      \IEEEauthorblockA{
        \hpcaaffiliation{} \\
        \hpcaemail{}
      }
  \else
    \IEEEauthorblockN{\normalsize{HPCA \hpcayear{} Submission
      \textbf{\#\hpcasubmissionnumber{}}} \\
      \IEEEauthorblockA{
        Confidential Draft \\
        Do NOT Distribute!!
      }
    }
  \fi 
}

\begin{document}
\maketitle

\ifdefined\hpcacameraready 
  \pagestyle{empty}
\else
  \thispagestyle{plain}
  \pagestyle{plain}
\fi

\newcommand{\hpcaheight}{0mm}
\ifdefined\eaopen
\renewcommand{\hpcaheight}{12mm}
\fi

\pagenumbering{arabic}

\definecolor{codegreen}{rgb}{0,0.6,0}
\definecolor{codegray}{rgb}{0.5,0.5,0.5}
\definecolor{codepurple}{rgb}{0.58,0,0.82}
\definecolor{backcolour}{rgb}{0.95,0.95,0.92}
\definecolor{LightBlue}{rgb}{0.83, 0.91, 1}
\definecolor{LightGreen}{rgb}{0.8, 1, 0.8}
\definecolor{LightPink}{rgb}{1, 0.8, 0.88}
\definecolor{LightYellow}{rgb}{1, 1, 0.6}

\lstdefinestyle{mystyle}{
    backgroundcolor=\color{white},   
    commentstyle=\color{codegray},
    keywordstyle=\textbf,
    numberstyle=\tiny\color{codegray},
    stringstyle=\color{codepurple},
    basicstyle=\ttfamily\fontsize{6}{6}\selectfont,
    breakatwhitespace=false,         
    breaklines=true,                 
    captionpos=t,                    
    keepspaces=true,                 
    numbers=none,                    
    numbersep=5pt,                  
    showspaces=false,
    showstringspaces=false,
    showtabs=false,        
    tabsize=2,
    frame=single,
    framesep=0pt
}

\lstset{style=mystyle}


\newif\ifrevision

\revisiontrue

\widowpenalty=1500
\clubpenalty=1500
\setlength{\emergencystretch}{3em}



\begin{abstract}
Bit-serial 
Processing-In-Memory (PIM) is an attractive paradigm for accelerator architectures, for parallel workloads such as Deep Learning (DL), because of its capability to achieve massive data parallelism at a low area overhead
and provide orders-of-magnitude data movement savings 
by moving computational resources closer to the data.
While many PIM architectures have been proposed, improvements are needed in communicating intermediate results to consumer kernels, 
for communication between tiles at scale, 
for reduction operations, and for  efficiently performing bit-serial operations with constants.


We present \sysname, a scalable architecture that provides spatially aware communication network for efficient intra-tile and inter-tile data movement and  provides efficient computation support for generally inefficient bit-serial compute patterns. 
Our architecture consists of a massive hierarchical array of compute-enabled SRAMs (CRAMs),
which is codesigned with a compiler to achieve high utilization.
The key novelties of our architecture are (1) in providing efficient support for spatially-aware communication by providing local H-tree network for reductions, by adding explicit hardware for shuffling operands, and by deploying systolic broadcasting, as well as (2) by taking advantage of the divisible nature of bit-serial computations through adaptive precision, bit-slicing and efficient handling of constant operations.
These innovations are integrated into a tensor expressions-based programming 
framework (including a compiler for easy programmability) that enables simple programmer control of optimizations for mapping programs into
massively parallel binaries for millions of PIM processing elements.

When compared against a similarly provisioned modern Tensor Core GPU (NVIDIA A100),
across common DL kernels and an end-to-end DL network (Resnet18),  \sysname outperforms the GPU by \gpuspeedup,
and reduces energy by \gpuenergy.
We compare \sysname with similarly provisioned state-of-the-art SRAM PIM (Duality Cache) and DRAM PIM (SIMDRAM), and observe a speedup of \dcspeedup and \simdramspeedup respectively.

\end{abstract}
\freefootnote{* Aman Arora and Jian Weng are co-first authors with equal contribution.}

\section{Introduction}

Bit-serial Processing-In-Memory (PIM) is a promising accelerator paradigm
\cite{neural_cache,duality_cache,floatpim,cim_stt_mram,drisa}
with both high compute density and abundant on-chip memory capacity,
especially considering the recent surge of demands
on computing power and memory bandwidth
in multiple application domains, including but not limited to
deep learning, image processing, and signal processing.
The essential principle of this paradigm is
to integrate a single-bit processing element (PE) at the output of the sense
amplifier under each bitline of a memory array so that
massive data parallelism can be exploited over a transposed data layout.


This technology provides compute density that is competitive with the state-of-the-art GPUs.
The theoretical throughput of
a PIM system based on prior technologies~\cite{comefa,neural_cache} is in the range
of 310-340 GOPS/mm$^2$ for int8 precision, for the same area
and DRAM bandwidth as that of an NVIDIA A100 GPU. The
GPU has a much lower vector throughput of 24 GOPS/mm$^2$, but has a higher throughput of 755 GOPS/mm$2$ for Tensor
Cores. However, Tensor Cores can only
achieve high utilization for specific kernels and parameters.
In addition, bit-serial PIM supports arbitrary precision, which
can be extremely beneficial for saving memory bandwidth and
increasing compute throughput. The paradigm keeps data near
compute units to avoid data movement overhead and thwart
the memory wall~\cite{mem-wall}. Overall, bit-serial PIM is a promising
paradigm that has competitive compute density
without needing specialized units like Tensor Cores, and can
be a path-forward for DL workloads. 

State-of-the-art PIM systems~\cite{duality_cache, simdram_asplos_2021} have showcased improved performance compared to previous generation GPUs.
To make PIM systems outperform the state-of-the-art GPUs, we need to fully unlock
the potential of the PIM paradigm by taking a system-level approach - co-optimizing  hardware and software.
Hardware should be carefully architected, given the area budget, to optimize computation and communication.
Prior works suffer from
excessively high overhead in on-chip data communication, because of lack of
hardware specialization for common data access behaviors.
Similarly, the software can be tuned to make better
use of the underlying hardware. 
Prior works do not enable the software to take advantage of the hardware's bit-serial nature to perform optimized data allocation and computation.
Also, though some prior works claim to have
a full-stack implementation, their programming
interfaces are rather low-level. These interfaces limit
the productivity of application development and performance tuning.

Our goal is to build a Processing-In-Memory (PIM) system - ISA, microarchitecture and compiler - that 
can exceed the performance
and energy efficiency of similarly-provisioned GPUs and prior PIM systems, with a focus on DL workloads.
There are two key principles that form the basis of our proposed design: 
1. We optimize on-chip \textbf{communications to be spatially-aware}:  \underline{H-tree interconnect} topology for faster reductions at lower level of hierarchy \& \underline{dynamic routing} at higher level of hierarchy for scalability, \underline{explicit hardware for shuffling \& multicasting} operands for common data patterns in DL workloads, and \underline{systolic broadcasting}.
2. We \textbf{optimize the bit-serial computations} that are common in PIM architectures: memory allocation can expand/shrink dynamically based on precision requirements (\underline{adaptive precision}), large precisions can be broken down into smaller parallel computations (\underline{bit-slicing}), and saving space \& time by exploiting bit-level sparsity in operands for \underline{constant operations}.
Operations such as reductions, constant multiplication, multicasting, broadcasting are very common in workloads like DL.

\ifrevision

\else
\begin{figure}
    \centering
    \includegraphics[width=\linewidth]{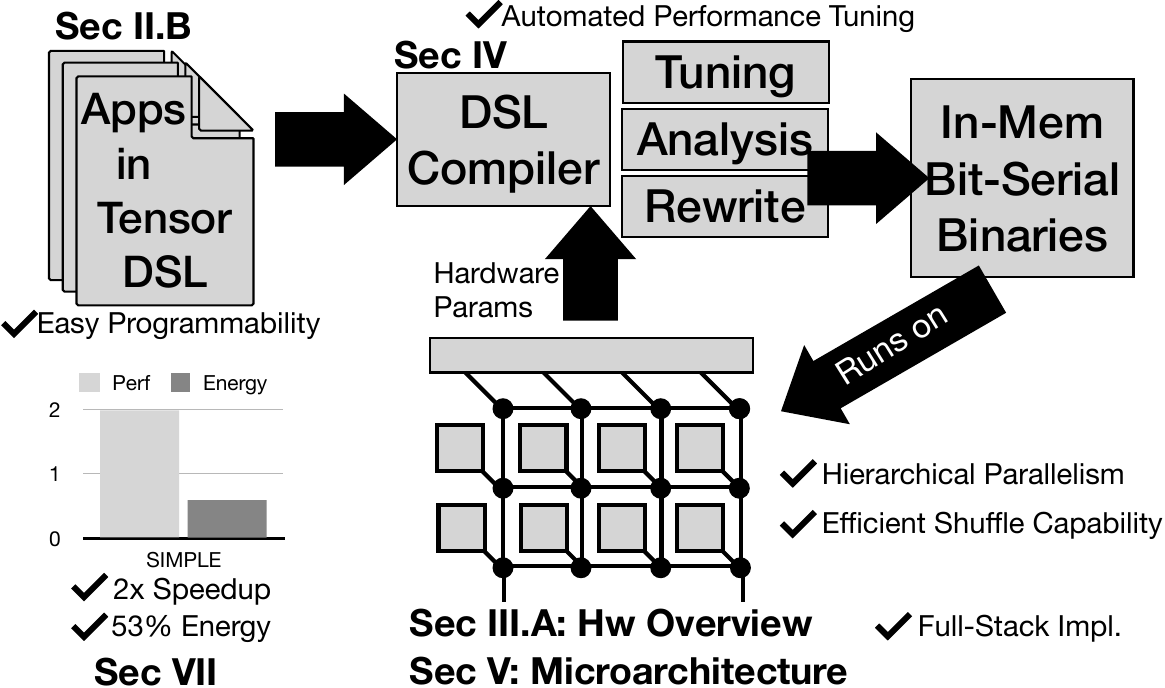}
    \caption{High-level overview of \sysname}
    \label{fig:sys-overview}
\end{figure}
\fi

Our overall system is a hierarchical and spatial PIM accelerator, abbreviated as \sysname.  
\sysname uses a hierarchical structure, where each tile
is composed of many SRAM arrays capable of bit-serial PIM, along
with an instruction controller that broadcasts commands to
SRAM arrays in its tile. The ISA enables efficient expression of mixed scalar/vector
program regions.  The intra-tile network is simple and static
for low overhead, and uses an H-tree \cite{interconnections_book} to facilitate high-bandwidth reduction. 
Broadcasting, multicasting, and shuffling mechanisms
can be configured on this H-tree to reduce data packing overheads.
At the inter-tile level, tiles communicate explicitly,
and routing is done over a dynamically routed network to 
enable flexible parallelization strategies. 
Further, a mesh-based topology enables scalability to arbitrary sizes.
\sysname's programming interface is based on the TVM tensor DSL~\cite{tvm},
which can be used to express a wide range of applications,
including linear algebra, neural networks, and stencil processing.
With moderate hints from the developers,
the compiler can easily generate portable and high-performance code,
by partitioning work across millions of
PEs and balancing buffer occupancy and data parallelism.

Our evaluation shows that with sufficient co-design,
\sysname can rival and surpass
state-of-the-art GPUs as well as prior PIM systems.
Specifically, we achieve \gpuspeedup speedup over NVIDIA A100,
while having \gpuenergy energy improvement 
for the equal-provisioned area and the same memory bandwidth.
We also observe a speedup of \dcspeedup with similarly provisioned state-of-the-art SRAM PIM (Duality Cache), and a speedup of \simdramspeedup with similarly provisioned state-of-the-art DRAM PIM (SIMDRAM).
To sum up, the contributions are:
\begin{itemize}
    \item A hierarchical and spatial PIM system with an ISA, a microarchitecture, a compiler and a programming interface.
    \begin{itemize}
        \item A microarchitecture that deploys dual-ported SRAM arrays with configurable PEs for PIM. 
        \item An ISA that exposes PIM-specific features of the hardware that can be utilized by the compiler. 
        \item A compiler that can automatically tune the parallelism and on-chip buffer allocation, with moderate hints from application developer.
        \item A user-friendly programming interface using TVM Tensor DSL.
    \end{itemize}
    \item Employing techniques for spatially-aware communication (shuffle hardware, H-tree for efficient reduction, systolic broadcasting) and bit-serial-aware computation (constant operations, adaptive precision, bit-slicing) to achieve high performance.
    \item Demonstration of GPU-outperforming performance and energy efficiency across both DL microbenchmarks and an end-to-end Deep Neural Network (DNN).
    \item Comparison with state-of-the-art SRAM and DRAM PIM systems, showing improved performance for realistic benchmarks.
\end{itemize}

\begin{figure}[t]
\centering
\includegraphics[width=0.9\linewidth]{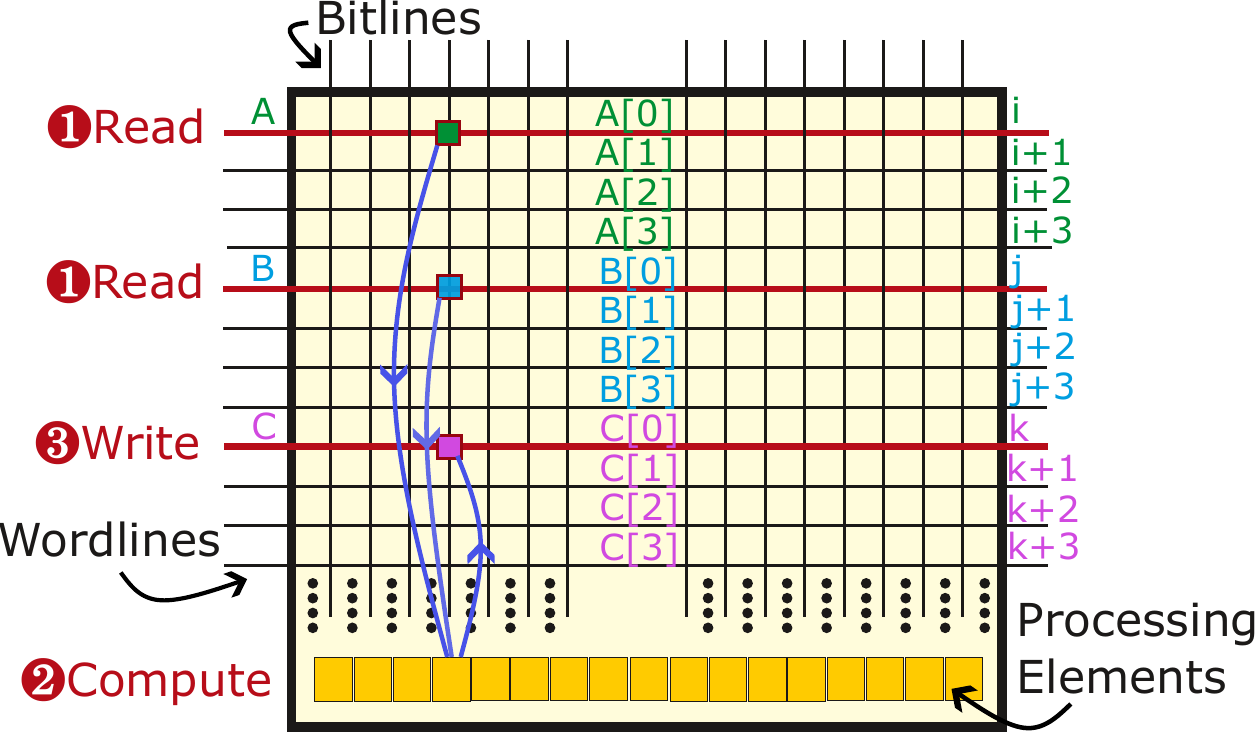}
\caption{Basics of bit-serial Processing-In-Memory}
\label{fig:cram_operation}
\end{figure}

\section{Background} \label{sec:bg}

\begin{figure}[t]
    \centering
    \includegraphics[width=0.8\linewidth]{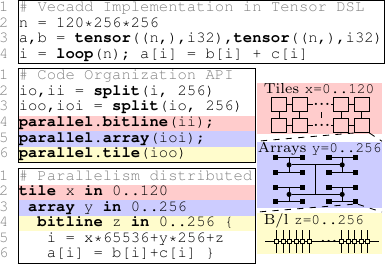}
    \caption{Programming \sysname in tensor DSL.}
    \label{fig:dsl-example}
\end{figure}

\begin{figure*}[t]
    \centering
    \includegraphics[width=0.9\linewidth]{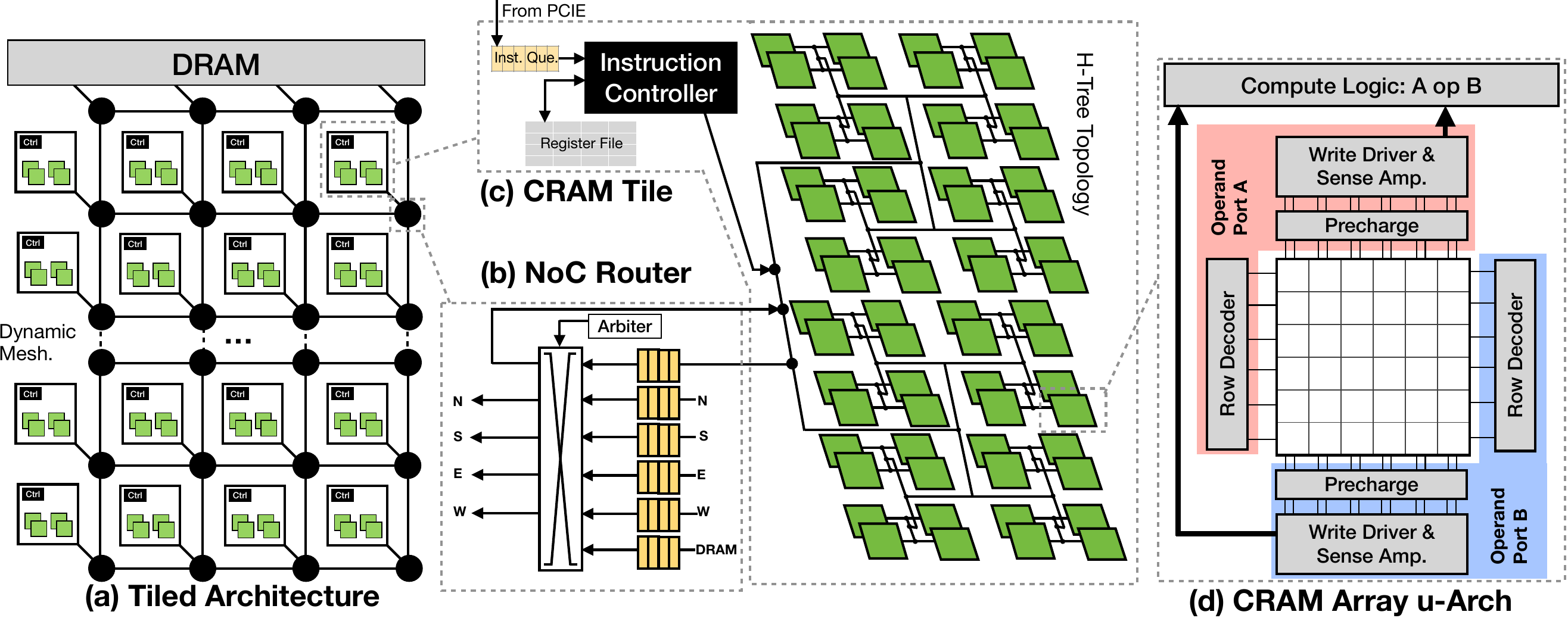}
    \caption{\sysname hardware architecture}
    \label{fig:arch-overview}
\end{figure*}

\subsection{Bit-serial Processing-In-Memory in SRAMs} \label{sec:bg-pim}
Bit-serial computing paradigm performs operations on data bit-by-bit instead of element-by-element. This makes each operation take many cycles, but massive parallelism can be achieved by utilizing simple 1-bit processing elements, enabling high throughput.

Bit-serial Processing-In-Memory combines bit-serial computing with Processing-In-Memory. 
\rev{
Analog approaches to bit-serial PIM \cite{shanbag_sram_pim_analog, shanbag_sram_pim_analog2} require analog-to-digital and digital-to-analog converters, have high power consumption, and are therefore, not considered in this work.
In digital approaches, 1-bit processing elements (PE) are added to an SRAM block. To provide operands to the PEs, two methods are used: 
(1) activating multiple wordlines at the same time on one port \cite{compute_cache,neural_cache,sram_pim_batten,duality_cache},
(2) using dual ported RAMs to read two wordlines at the same time \cite{comefa}.
Table \ref{table:diff_neural_cache_comefa} compares various properties of the compute capable SRAM blocks that use these two methods.
}

\begin{table}[t]
  \renewcommand{\arraystretch}{1.1}
  \centering
  \caption{\rev{Single-port based vs. dual-port based SRAM PIM}}
  \label{table:diff_neural_cache_comefa}
  \footnotesize
  \begin{tabular}{
  | >{\raggedright\arraybackslash}m{0.48\columnwidth} 
  | >{\centering\arraybackslash}m{0.16\columnwidth}
  | >{\centering\arraybackslash}m{0.16\columnwidth}  
  |}
    \hline
    \rowcolor{LightBlue} \centering
    \textbf{Feature} & \textbf{Single Port Based} & \textbf{Dual Port Based}\\
    \hline

Activate two wordlines at the same time on one port    & Yes   & No   \\ \hline
Requires extra voltage source & Yes   & No \\ \hline
Requires extra row decoder    & Yes   & No \\ \hline
Requires modification to sense amps    & Yes   & No \\ \hline
Compute uses dual-port behavior     & No    & Yes \\ \hline
Generic/Flexible PE    & No    & Yes \\ \hline
Cross-RAM shift    & No    & Yes \\ \hline
Area of the array & Lower & Higher \\ \hline
Frequency of operation & Lower & Higher \\ \hline 
Examples & \cite{compute_cache,neural_cache,sram_pim_batten,duality_cache} & \cite{comefa} \\ \hline

    \hline
  \end{tabular}  
\end{table}


Figure \ref{fig:cram_operation} shows the basic principle of bit-serial Processing-In-Memory.
\ifrevision
In every cycle, two wordlines containing a bit of each operand are activated, the processing element performs the computation and the result is written into a wordline.
Operations such as addition, multiplication, etc. can be performed by repeating this basic step over multiple cycles. We refer the reader to Neural Cache \cite{neural_cache} for a detailed description of the algorithms for various operations. Note that floating point and transcendental operations are also supported \cite{floatpim,duality_cache}.

The challenges in prior PIM systems include (1) high on-chip communication overhead in moving partial results across the chip and in organizing the data in the right layout, especially in large systems with thousands of RAMs, even though the off-chip memory traffic is reduced, and (2) bit-serial computation takes a large number of cycles, especially when the precision expands because compilers allocate the number of bits based on traditional paradigms (e.g. int8 $*$ int8 $->$ int32) .
\else
\fi

\ifrevision
\else
Consider an example of bit-wise \texttt{AND}ing of the elements of two arrays (array length = number of bitlines and element width = 4 bits).  Each element is stored in a column (bitline), 1 bit in 1 row (wordline). Elements of array 1 are stored in rows $i$, $i+1$, $i+2$ and $i+3$. Elements of array 2 are stored in rows $j$, $j+1$, $j+2$ and $j+3$. A total of 8 rows are required to store both arrays. In one cycle, (1) rows $i$ and $j$ are read on two ports, (2) the processing element computes the \texttt{AND} of two bits 
and (3) the result is stored in row $k$. This process is repeated 4 times with increasing row addresses, and the final result is available, after 4 cycles, in rows $k$, $k+1$, $k+2$ and $k+3$. 
Note that a transposed data layout is used and that many operations are done in parallel in each cycle, making the RAM a SIMD 
engine.

Arithmetic operations follow a similar data layout and operation mechanism. For performing \textbf{addition} operation, in each cycle, the processing element computes the sum, stores the carry for the next cycle computation, and writes back the sum bit to the result row. Then the next bit position is processed in the next cycle, using the stored carry. The final carry is stored into a row using an extra cycle. Thus, the addition for $n$-bit operands takes $n+1$ cycles. \textbf{Multiplication} is based on iterative addition of partial results. In each iteration, one bit of the first operand is loaded as into a \texttt{mask} flip-flop, and the second operand’s bits are added to the partial sum only if \texttt{mask} is 1. Multiplication of $n$-bit operands takes $n^2 + 3n -2$ cycles.
The \textbf{reduction} operation is implemented as a sequence of across-bitline shifts and additions. In each step, the number of operands becomes half, until the final result is generated in one bitline. 
Note that floating point and transcendental operations are also supported \cite{floatpim,duality_cache}.
\fi




\subsection{Tensor Domain-Specific Language (DSLs)}

DSLs, like Halide~\cite{halide}, 
TVM~\cite{tvm}, and Tensor Comprehensions~\cite{tensor-comp},
are developed to productively write
high performance tensor programs.
The idea is to decouple the algorithm and the performance tuning
controlled by loop re-organization.
Consider the vector addition implementation in Figure~\ref{fig:dsl-example}:
Loop variables and tensors are first declared,
and then a vector addition is implemented
in an expression involving these declared variables.
Using the DSL allows us to tune the algorithm at an abstract level  orthogonally with the specific problem tiling and work distribution that is involved when realizing it on a specific hardware.
By tiling, ordering, and annotating the loops,
the parallelism in the program can be mapped onto our hardware hierarchies.

\section{Overview} \label{sec:overview}

\ifrevision
\else
Our goal is to build a scalable system that effectively exploits massive data parallelism within applications and is easy to program.
In this section, we
provide an overview of the hardware organization and describe the ISA of the system.
\fi

\subsection{Hardware Organization}

Figure~\ref{fig:arch-overview} provides an overview of the hardware organization of \sysname.
The \sysname hardware deploys a large number of compute-enabled SRAMs (or \textbf{CRAMs}). 
Each CRAM is a dual-ported SRAM modified to add multiple single-bit PEs. We base our CRAM design on CoMeFa \cite{comefa}, \rev{because of its more practical design compared to Neural Cache \cite{neural_cache}}. 
We use their basic block to build a large scalable network of CRAMs with several enhancements for both communication and computation, enabling  efficiently at scale. 

To communicate between the CRAMs, a \textbf{statically scheduled network} is chosen, since most communication patterns are identifiable at compile time. 
We choose an \textbf{H-Tree} topology for this network, \rev{because it is
well suited for partial sum reduction, a common computation pattern in  DL and many modern  applications}.
Statically scheduling the entire chip would put too much burden on the compiler. So, we introduce another level of hierarchy: \textbf{tiles}. Tiles communicate using a dynamically scheduled packet-switched \textbf{NoC}. We choose a \textbf{2D mesh} topology for the NoC because this enables scalability. The NoC is used to send and receive data across tiles and to/from DRAM.
Having parallelism at 3 levels of hierarchy - CRAM, tile, chip - enables \sysname to capture different types of parallelism
in highly data-parallel applications.

Each CRAM needs to be fed micro-ops to perform computation. An \textbf{instruction controller} decodes the instructions and provides micro-ops to the CRAMs every cycle. However, connecting an instruction controller to each CRAM would result in significant overhead. We reduce this overhead by having one instruction controller in a tile, making CRAMs in each tile operate in a SIMD fashion. 

The \sysname system defines three memory locations: main memory (\textbf{DRAM}), CRAMs and \textbf{register file}.
HBM (High Bandwidth Memory) DRAM is adopted to sustain the high bandwidth
required by massive parallelism.
To simplify physical design,
DRAM controllers are connected to the routers at the edges of the NoC. For similar reasons, we connect all DRAM controllers to the top edge of the mesh NoC.
A register file is provided in each tile to store constants or scalars.
We assume a PCIe interface, both for loading
instructions and for transferring data (like GPUs).

\subsection{Enabling Scalable and Performant Processing-In-Memory}

In this section, we present the innovationse - for spatially-aware communication and for bit-serially-aware computation - that make \sysname a scalable and performant PIM system.

\noindent\textbf{Register File and Constant Operations}:
A frequent operation in applications such as DL is multiplying a scalar (or a constant) with an array or vector operand. With the computation paradigm of bit-serial PIM, we would have to replicate this scalar over multiple bitlines in the CRAM. A more efficient way is to keep scalars outside the CRAM and perform what we call constant multiplication (explained in Section \ref{sec:uarch}). 
To store these scalar operands, we introduce a \textbf{register file (RF)} in each tile.
Additionally, this approach can enable exploiting bit-level sparsity in the
constant operand by skipping operations for zero-bits, leading to up to 2$\times$ speedup in operations like multiplication and 4$\times$ speedup in operations like dot product.
\rev{This feature is exposed to the compiler through the ISA (\texttt{mul\_const} instruction).}

\noindent\textbf{Dedicated Shuffle Hardware}:
When data is loaded from the DRAM, it often needs to be broadcasted or multicasted to various CRAMs in different patterns to avoid loading data multiple times or to ensure high utilization of CRAMs in a tile. 
In addition to just loading data from DRAM, broadcasting or multicasting is useful when data is transferred from one tile to another.
We provide explicit hardware near each CRAM to support this. 
Several multicast and broadcast patterns are supported, governed by the requirements of common workloads such as GEMM.
\rev{Pattern specification is exposed to the compiler through special fields in the data transfer instructions.}

\noindent\textbf{Adaptive Precision}:
Since \sysname uses bit-serial operations, any precision is supported, including floating point. In \sysname ISA, the precision for each operand can be specified separately. This capability of specifying a custom precision at operand granularity enables using just the number of bits that are required and allocating only the required number of wordlines.  For example, when multiplying numbers of precision 8 and 10, 18 wordlines can be allocated to store the result, instead of 32 bits as in a normal CPU. \rev{Our compiler exploits this feature to pack as any operands as possible in each bitline (enabling high reuse), even splitting portions of an operand across non-consecutive wordlines.}

\noindent\textbf{Bit Slicing}:
Although bit-serial PIM can support any precision, the number of cycles consumed by an operation directly depends on the precision of the operands. Larger precisions (e.g. \texttt{int32}) take more cycles than smaller precisions (e.g. \texttt{int8}).
However, in \sysname, we employ a technique where larger precision operations are broken down into smaller-precision independent parallely-executable operations and the results are combined later. We refer to this as bit slicing. 
\rev{For addition, this is exposed to the compiler by a special field in the instruction. For multiply, this is done by the compiler in software.}

\noindent\textbf{Cross-CRAM Shift}:
Shifts are commonly used in operations like stencils, filters, etc. 
Vectorization widths in PIM architectures can get really large (e.g. in \sysname, the vectorization width for maximum utilization is 256*256). 
Supporting only intra-CRAM shifting (i.e. shifting data from a bitline to the next within a CRAM using connections between PEs) limits the utility of the shift operation to only a CRAM.
To support shifting data from a bitline to the next across the whole vectorization width, we provide CRAM-to-CRAM shift connections within a tile.
This gives \sysname the ability to perform filters and stencils much more efficiently.

\noindent\textbf{Systolic Broadcasting}:
Chip level communications, such as broadcast, are essential for workloads such as convolutions, where weights need to be broadcasted to multiple compute units or tiles.
However, naive broadcast algorithms, like one-to-many transfers, can cause extreme network congestion and overheads.
\rev{To optimize this, we support a near-neighbor systolic-like data transfer supported in hardware and exposed to the compiler through the \texttt{load\_bcast} and \texttt{tile\_bcast} instructions.
This efficiently utilizes the available NoC bandwidth and reduces congestion.}

\noindent\textbf{Hierarchical Interconnect}: A two-tiered interconnect is used in \sysname. A statically scheduled H-tree interconnect topology at the lower (intra-tile) level is used to enable faster reductions. A dynamically scheduled mesh interconnect topology at the higher (inter-tile) level is used to reduce the burden on the compiler and also to increase the scalability of \sysname.
\rev{
A non-hierarchical \sysname would mean one of two implementations: 
(1) All CRAMs connected using a static H-tree interconnect. This would lead to excessive burden on the compiler because the whole computation would need to be statically scheduled. Additionally, this would cause a rapid degradation in the frequency of operation of the chip as the number of CRAMs increases.
(2) All CRAMs connected using a dynamically routed mesh network. This would imply one router for each CRAM, increasing the area overhead of non-computational logic significantly. 
In either implementation, we would need one instruction controller per CRAM.
Area overheads of each unit are shown later in Section \ref{section:area_results}.
}

\section{Architecture} \label{sec:arch}

\subsection{Instruction Set Architecture (ISA)} \label{section:isa}

In this section, we elaborate the \sysname ISA, including
Compute, Data Transfer and Synchronization instructions.




\noindent\textbf{Compute Instructions}: 
Compute instructions support arithmetic and logical operations, operate on data in the CRAMs, and are vectorized across bitlines.
We also support inter-bitline instructions, like shifting data across bitlines.
Instructions to reduce data within a CRAM and across the CRAMs in a tile are also provided.
We also have an instruction, \texttt{set\_mask}, which copies the data of wordline into the mask latches in PEs, to enable predicating operations per bitline.
Additionally, each instruction has a field to specify what should be used for predication - the mask latch or the carry latch (Section \ref{sec:uarch} describes the PE architecture incl. the latches).
In most cases, all compute instructions are executed across all the
CRAMs in tile, but we also have a field (called \texttt{size}) to specify the number of bitlines involved in the operation across the tile.




\noindent\textbf{Bit Slicing}:
To support bit slicing, in the \texttt{add} instruction, we provide two special fields.
The \texttt{cen} field can be used to specify if the carry stored in the PE (from a previous op) should be used during the first step. The \texttt{cst} field can be used to specify if the generated carry in the last step should be written into the CRAM. This allows larger precision addition operations to be broken down into smaller precision operations by the compiler much more efficiently. For example, to perform an 8-bit addition, two separate 4-bit additions can be performed, and the second 4-bit addition can directly use the CARRY bit from the first one.
Also, for multiplication ops, bit slicing is done at the software level by the compiler, which divides-up large operands, performing the individual ops in parallel and reducing the results.





\noindent\textbf{Operations with scalars or constants}:
For multiplication operation, a special instruction called \texttt{mul\_const} is provided where one operand is from the RF (scalar or constant), instead of being replicated in the CRAM.
This instruction skips zeros in the constant operand in the RF, reducing the
execution time.


\noindent\textbf{Data transfer instructions}: These instructions
are used to move data between the DRAM, CRAMs and the RF.
Specifically, we support bidirectional data transfer
between DRAM and CRAMs, as well as DRAM and the RF.
An instruction to send data from one CRAM in a tile to another is also provided.
In addition, we also support data loaded from DRAM to be broadcast to
multiple tiles' CRAMs. Within each tile, we allow one CRAM
to broadcast data to all other CRAMs.
Tiles are allowed to perform either point-to-point communication
or broadcast data. Point-to-point communication blocks
the receiver until the data arrives.

\noindent\textbf{Synchronization instructions}: These instructions coordinate data transfers and computations among tiles.
Two synchronization instructions provided are \texttt{signal} and \texttt{wait}. 
\texttt{signal} sends a message from a source CRAM to a destination CRAM and is non-blocking.
A CRAM can wait for a message (blocking) from a source CRAM using the \texttt{wait} instruction.

\noindent\textbf{Transposing data}:
In load and store instructions, besides the source address, destination address, size and  precision, there is an additional \texttt{tr} field specifying if the data is transposed or not. This can be used when, e.g., an immediate/constant operand read from the main memory need not be transposed.

\noindent\textbf{Program example}:
A simple elementwise vector multiplication is shown in Listing \ref{listing:simple_mult_pgm}. The program generates an instruction stream for all tiles in the chip (\texttt{NUM\_TILES}). 
Two operand arrays, each with elements of precision \texttt{int8}, are loaded from the main memory. \texttt{vec\_width} is specified to be the full width of a tile. 
Then a multiplication instruction is used to generate a result with precision \texttt{int6}. 
The result is then stored back to main memory. 

\begin{lstlisting}[label=listing:simple_mult_pgm, caption=Simple program to add two arrays, basicstyle=\scriptsize]
int vec_width = NUM_CRAMS_IN_TILE * NUM_BITLINES_IN_CRAM;
for (i = 1; i<NUM_TILES; i++) {
  load tile_addr1, dram_addr1, vec_width, i8
  load tile_addr2, dram_addr2, vec_width, i8
  mult tile_addr3, i6, tile_addr2, i8, tile_addr1, i8
  store dram_addr3, tile_addr3, vec_width, i16 }
\end{lstlisting}

\subsection{Microarchitecture} \label{sec:uarch}

Here we discuss \sysname's microarchitecture. 
Table \ref{tab:hardware_params} provides a list of hardware
parameters.

\begin{table}[bh]
\centering
\caption{Microarchitectural parameters of \sysname}
\footnotesize
\begin{tabular}{|l|l|}
\hline
\rowcolor{LightBlue} \centering
\textbf{Parameter} & \textbf{Value} \\
\hline
CRAM geometry & 256x256 \\
\hline
PEs per CRAM & 256 \\
\hline
CRAM size & 8 KB \\
\hline
Mesh dimensions & 12x10 \\
\hline
DRAM bandwidth & 12288 bits/clock \\
\hline
Clock frequency & 1.5 GHz \\
\hline
Num tiles & 120 \\
\hline
Num CRAMs per tile & 256 \\
\hline
Total CRAMs & 30720 \\
\hline
RF size & 32 32-bit regs \\
\hline
Tile-to-Tile bandwidth & 1024 bits/clock \\
\hline
CRAM-to-CRAM bandwidth & 256 bits/clock \\
\hline
\end{tabular}%
\label{tab:hardware_params}
\end{table}

\noindent\textbf{CRAMs:}
We employ dual-ported compute-enabled RAMs (called CRAMs).  
A CRAM has two modes: compute and memory. In compute mode, the data word written into the memory is treated as a micro-op. Each micro-op takes 1 cycle, in which two wordlines are read, computation is performed in the PE, and the result is written into a wordline. In memory mode, the CRAM behaves as a normal RAM.
CRAMs are grouped into tiles; all CRAMs in a tile execute in lock-step in a SIMD fashion (except when executing CRAM-to-CRAM data transfer). 
CRAMs in a tile are connected through the intra-tile network. In addition, there is a single wire ring interconnect between all CRAMs in a tile to facilitate \texttt{shift} instructions. 

\begin{figure}[t]
    \centering
    \includegraphics[width=0.95\linewidth]{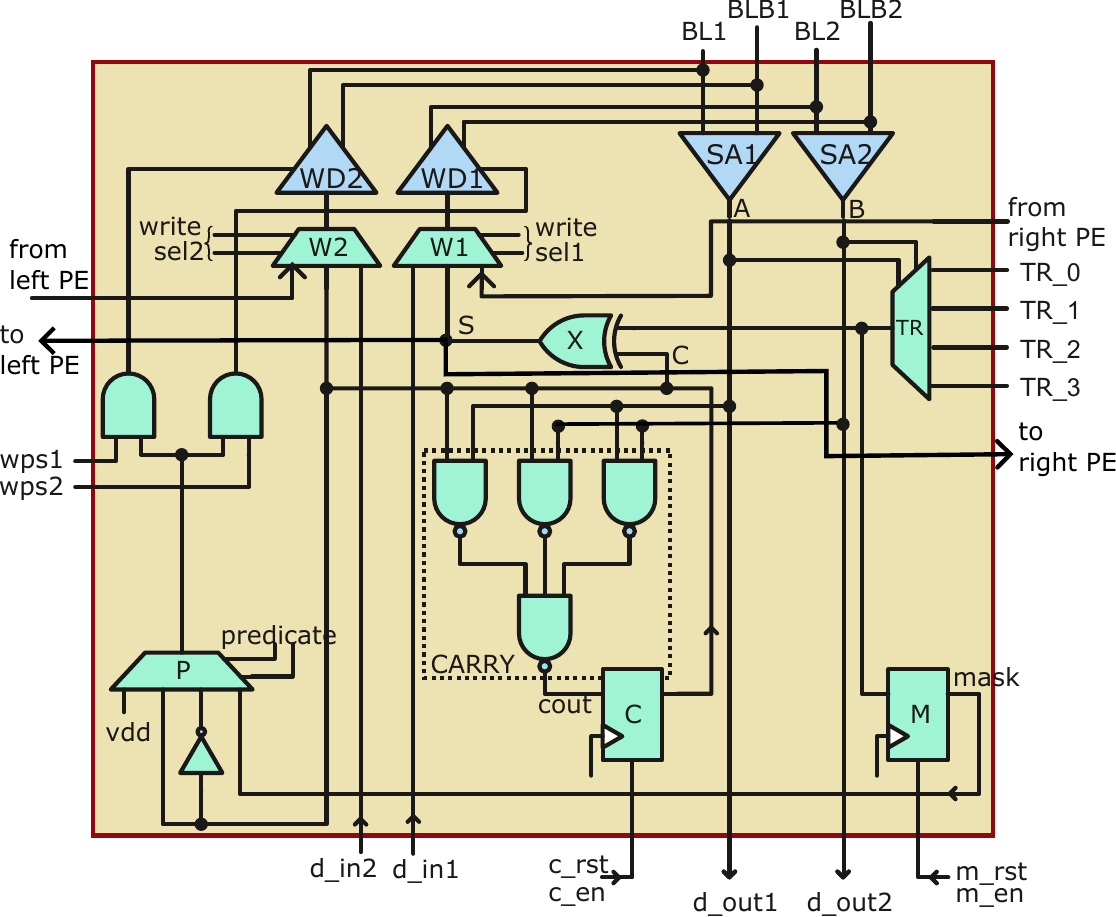}
    \caption{Processing element used in CRAMs}
    \label{fig:processing_element}

\end{figure}

\noindent\textbf{Processing Element (PE):}
\sysname adopts the PE architecture from  CoMeFa \cite{comefa},
 as shown in Figure \ref{fig:processing_element}. 
Each PE can perform any logical operation between 2 operands, using the \textbf{TR} mux.
With the addition of an XOR gate (\textbf{X}), it can also perform a 1-bit full adder operation. A carry latch (\textbf{C}) is used to store the carry-out, which can be used as carry-in for the next timestep.
The output of the TR mux can be stored in the mask latch (\textbf{M}).
Predication based on mask bits and carry bits is supported, through the predication mux (\textbf{P}).
There are as many PEs in a CRAM as many bitlines.
The operation performed by the element is governed by the micro-op received by the CRAM from the instruction controller.

\noindent\textbf{Instruction controller:}
Instructions are received from the HOST over PCIe. Each tile has an instruction controller to decode and farm-out execution to corresponding units. 
For compute instructions (add, multiply, reduction, etc.), it generates micro-ops for the CRAMs every cycle.
For data transfer instructions (CRAM-to-CRAM transfer, tile-to-tile transfer, DRAM transfers), it reads the CRAM and sends data into the static network's switches, and also writes data coming in from the switches into the CRAMs.

\noindent\textbf{On-chip networks:}
The inter-tile network uses a standard wormhole-switched dynamic NoC, with X-Y routing.
The intra-tile network is a static circuit-switched network using an H-Tree topology. This is similar to a hierarchical FPGA \cite{hierarchical_fpga,hsra}, but with much smaller configuration overhead because of the coarser granularity (word-level instead of bit-level).
Each switch is a buffered crossbar with 5 input and output ports. Each output port can be driven by the other 4 input ports, controlled with 2 configuration bits. 
This network only needs to be reconfigured if there is a data transfer instructions with new communication pattern; this is
rare for tile-to-tile transfers and DRAM transfers.


\noindent\textbf{Shuffle logic:} To facilitate operations like GEMM and convolutions, different data layout patterns are required. E.g. we may need a data element to be duplicated in each bitline or repeated every 4 bitlines in a CRAM. Such data layouts help data reuse and avoid unnecessary data traffic from/to DRAM.
For this purpose, at the periphery of each CRAM, a shuffle logic unit is provided. Data can be shuffled before writing into the CRAM.
The shuffling pattern is an input to this shuffle logic and is exposed to the compiler through the \texttt{shf} argument in the \texttt{tile\_bcast} and \texttt{load\_bcast} instructions. 
\texttt{shf} specifies the stride of distributing the bits of data received onto the CRAMs in the tile.
E.g., a 256-bit value loaded from DRAM into a tile can be shuffled such that the first bit is duplicated in 256 bitlines in the first CRAM, the second bit is duplicated in 256 bitlines in the second CRAM, and so on.


\noindent\textbf{DRAM interface and transpose unit:}
All tiles in the top-row of the mesh NoC are connected to DRAM controllers.  
The data from DRAM must be transposed before storing into CRAMs, so that bit-serial arithmetic can be performed. Results need to be untransposed when writing back.
We use a transpose unit similar to CoMeFa's \cite{comefa}. 
\rev{It employs a ping-pong FIFO. Data enters from one side into the ping part in the non-transposed format.
When full, transposed output is obtained by reading bit slices of the loaded elements, while the pong part is filled with new data.
When the pong part is full, the roles are reversed and the process repeats.
In \sysname, this unit is integrated within the DRAM controller(s).
This unit can be disabled if not needed (through the \texttt{tr} field of the DRAM load/store instructions), e.g. for the weights
of a neural network, which can be pre-transposed.}

\noindent\textbf{Register File and operations with constants}:
Each tile has a register file (RF) that stores
data in untransposed format. 
RF is used in performing operations with scalars or constants.
When an instruction (mul\_const and add\_const) is issued, the instruction controller fetches the scalar operand from the RF and sends micro-ops to the CRAMs according to the bits of the constant which are set.
This saves the overhead of copying the same operand to multiple bitlines in the CRAM. It also speeds up the computation by exploiting bit-level sparsity through skipping some micro-ops. For example, for multiplication, if a bit of the scalar operand is 0, all micro-ops for summing the partial products for that bit position can be skipped.
Finally, we enable parallel RF writes 
for quickly loading constants.
Thus, we use a flip-flop based design to avoid port restrictions.

\section{Compiler} \label{sec:compiler}

\subsection{Overview}

\begin{figure}[t]
    \centering
    \includegraphics[width=1\linewidth]{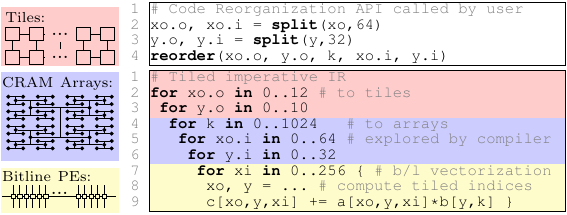}
    \caption{Reorganize the loops for parallelism distribution.}
    \label{fig:mv-distro}
\end{figure}

\paragraph{Programming Interface} \rev{We adopt a tensor DSL
as our high-level interface, because of its portability
and ease of performance tuning.
As shown in Figure~\ref{fig:mv-impl}(a),
a matrix multiplication is implemented
in a tensor DSL, by declaring loops, and tensors,
and describing the program behaviors
in expressions involving these declared variables.
The loop organizations are the key to the performance
tuning in tensor programs, and can be easily explored
by invoking several loop organization primitives (e.g.
\texttt{split} and \texttt{reorder} shown in Figure~\ref{fig:mv-distro}).
The parallelism is naturally encoded in the
declared loops with different types, either data-parallel or reduction.
These different loop types may lead to different program behaviors
when mapping loops to different hardware hierarchies.}

\begin{figure}
\hspace{-0.05in}
\begin{minipage}{0.54\linewidth}
\begin{lstlisting}[language=Python,
  emph={i32,i8,tensor,loop,red_loop}, emphstyle=\textbf]
# (a) Vanilla Matrix Multiply
n,m,p = 12*256*64, 10*32, 1024
a = tensor((n, p), i8)
b = tensor((m, p), i8)
x, y = loop(0, n), loop(0, m)
k = red_loop(0, p)
c[x,y] =
    sum(i32(a[x,k])*i32(b[y,k]))
\end{lstlisting}
\end{minipage}
\hfill
\begin{minipage}{0.45\linewidth}
\begin{lstlisting}[language=Python]
# (a') Imperative IR
for x in 0..n
 for y in 0..m {
  c[x,y] = i32(0)
  for k in 0..p
   c[x,y] +=
     a[x,k]*b[y,k]
}
\end{lstlisting}
\end{minipage}
\begin{minipage}{0.54\linewidth}
\begin{lstlisting}[language=Python,
  emph={i32,i8,tensor,loop,red_loop}, emphstyle=\textbf]
# (b) Relayout Matrix Multiply
a = tensor((n/256, p, 256), i8)
b = tensor((m, p), i8)
xo, xi = loop(0, n/256), loop(0, 256)
y = loop(0, m)
k = red_loop(0, p)
c[xo,y,xi] = sum(
  i32(a[xo,k,xi])*i32(b[y,k]))
\end{lstlisting}
\end{minipage}
\hfill
\begin{minipage}{0.45\linewidth}
\begin{lstlisting}[language=Python]
# (b') Imperative IR
for xo in 0..n/256
 for y in 0..m {
  c[xo,y,0..256] = i32x256(0)
  for k in 0..p
   # xi "vectorized" 0..256
   c[xo,y,0..256] +=
    a[xo,k,0..256]*b[y,k] }
\end{lstlisting}
\end{minipage}
    \caption{Matrix-matrix multiplication implemented in tensor expression language and array packing.}
    \label{fig:mv-impl}
\end{figure}

\paragraph{Performance Tuning}
Different implementations significantly affect the
on-chip buffer occupancy, memory traffic,
on-chip network traffic, and parallelism
distribution, and lead to different performances.
Considering the excessively large space of code organizations,
we decide to leave part of the performance tuning
as developers' responsibility, including the
loop organization and data layout, so that
the compiler can figure out the best parallelism
distribution and buffer allocation under this organization.

To explain, consider the matrix multiplication example in Figure~{\ref{fig:mv-impl}}(a).
Its imperative version in Figure~\ref{fig:mv-impl}(a')
shows that the innermost reduction loop across is hard to
be parallelized across bitlines specialized for vector parallelism.
Thus, one important transformation is to place
a data-parallel dimension in the inner loop.
As shown in Figure~\ref{fig:mv-impl}(b)\&(b'),
the outermost dimension of tensor \texttt{a}
is tiled by 256, and reordered to the innermost
for mapping to bitline PEs.
Then, Figure~\ref{fig:mv-distro} shows that users are required
to call the loop organization APIs to determine a code
organization for the compiler to distribute the parallelism
and allocate CRAM memory buffers.

\paragraph{Compiler Optimizations}
After the data layout and loop organization are determined,
the compiler analyzes the program and optimizes it.
The optimization includes both coarse grain optimizations
(in Section \ref{sec:compiler-coarse})
like distributing parallelism to hardware hierarchies,
and memory buffer allocation, as well as fine grain optimizations
(in Section \ref{sec:compiler-fine})
that take advantage of the properties of bit-serial arithmetic
to save on-chip memory occupancy.
Since the exploration space of coarse-grain optimizations is small,
the compiler exhaustively evaluates each point, and adopts the one with the best objective.
As shown in Figure~\ref{fig:mv-alloc}, the parallelism
distribution will affect the memory buffer allocation.
If the required buffer size exceeds the on-chip resources available,
this exploration point is considered invalid. To make more exploration
points more likely to succeed, fine-grain optimizations will
squeeze the buffer size requirement. 

\paragraph{Code Generation \& Feedback Loop}
After the favored parallelism distribution and buffer allocation
is \emph{successfully} determined on the given loop organization,
the compiler extracts all the computational instructions to be offloaded to
PIM and rewrites them in hardware intrinsics.
Then the transformed IR is ready for code generation.
If all the parallelism distribution fails under the given loop
organization, the compiler will throw an error to the developer,
and the developer is required to find another
more conservative loop organization.

\subsection{Parallelism Distribution \& Memory Allocation} \label{sec:compiler-coarse}
Parallelism distribution determines how much of these loops
should be tiled and parallelized across hardware hierarchies,
and how much should be executed in serial.
Since the parallel degree of each hierarchy is at an order of hundreds,
the loop tiling space is small enough for the compiler to search exhaustively.
Next, we explain how the loops are mapped to parallelism
across and within tiles.

\paragraph{Inter-Tile Parallelism Distribution} Considering the overhead
of communicating data between tiles, it is often inefficient to
reduce the partial sum across different tiles. Therefore, our compiler
only seeks to map data parallel loops to inter-tile parallelism.
Assuming we have 120 tiles, each iteration of \texttt{xo.o}
and \texttt{y.outer} in Figure~\ref{fig:mv-distro} are
mapped to each tile exactly. If the iterations exceed
the number of tiles, the compiler will seek to tile the loops
and execute parts serially.

\begin{figure}
    \centering
    \includegraphics[width=1.0\linewidth]{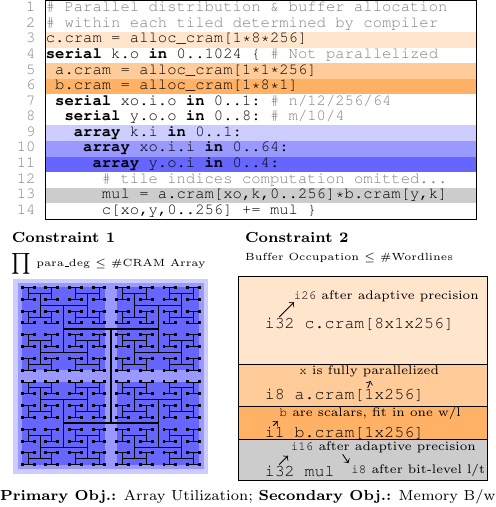}
    \caption{Distributing the parallelism intra-tile.}
    \label{fig:mv-alloc}
\end{figure}

\paragraph{Intra-Tile Parallelism Distribution}
Figure~\ref{fig:mv-alloc} shows that,
after the inter-tile parallelism is fixed,
the compiler distributes the intra-tile parallelism
by exploring the space of loop tiling.
Each tiled outer loop (with \texttt{.o} suffix)
will be executed in serial by each tile's controller,
and each tiled inner loop (with \texttt{.i} suffix)
will be mapped to a CRAM array. The total iterations of these
\texttt{.i}-loop should not exceed the number of arrays.
This can easily be enforced when tiling the loop by
multiplying the tiling factor. Besides, the CRAM buffer
allocation should not exceed the wordlines available
in each array.

When there are multiple distributions
that fulfill these two constraints (parallelization degree and CRAM buffer),
we use two objectives to determine the best one.
The primary objective is more computing resource occupancy,
and the second is less DRAM memory bandwidth. The rationale behind this objective order is that 
a high computing resource occupancy often requires high data bandwidth to sustain.

\paragraph{CRAM Buffer Allocation}
CRAM buffer allocation is the key to determine the feasibility of
a parallel distribution. Here, we first explain how the compiler greedily
exploits data reuse, and compute occupancy, while not exceeding the CRAM capacity. 
Some overused CRAM capacity can be false positive, the compiler will try
to detect and optimize it. 
If it turns out to be a true overuse, a feedback will be sent
to the developer for a conservative initial loop organization.

For the example shown in Figure~\ref{fig:mv-alloc},
the compiler greedily allocates the memory buffer at the highest 
serialized loop with reuse. Therefore, \texttt{a}
and \texttt{b} are allocated below \texttt{k.o},
and \texttt{c} is allocated above \texttt{k.o}.
Then, the compiler tries to minimize the CRAM buffer
occupancy by analyzing the data access pattern of each operand.
In the case shown in Figure~\ref{fig:mv-alloc},
the size of each buffer is proportional
to the iterations of serialized loops.
For example, because the indices of \texttt{c} are controlled by \texttt{xo} and \texttt{y},
the \texttt{c.cram} buffer size is \texttt{1}$\times$\texttt{8}$\times$\texttt{32},
where \texttt{1} is the serial iteration of \texttt{xo.i.o}, \texttt{8} is the
serial iteration of \texttt{y.o.o}, and \texttt{32} is the precision of the integers.
Similar thing happens to \texttt{a.cram}: because its index
is only controlled by \texttt{xo} (\texttt{k} is ignored, because there is no reuse over \texttt{k}), its buffer size is \texttt{1}$\times$\texttt{8}.
Therefore, parallelizing \texttt{xo.i} is more favored, because it save more
buffer occupancy for both \texttt{a}\&\texttt{c}, considering \texttt{b.cram} are all scalars.
After fully parallelizing \texttt{xo.i}, \texttt{y.o}
is further parallelized to fill the remaining arrays for compute resource occupancy.
Finally, \texttt{c}'buffer, 1x8x32=256 wordlines, already occupies each entire CRAM array,
with no space remaining for other operands, intermediate values, which indicates
the unfeasibility. However, it is a false positive. In the next section,
we will explain how we squeeze the CRAM allocation to optimize this false overuse.

\subsection{Optimizing CRAM Data} \label{sec:compiler-fine}

Here, we focus on optimizations within each CRAM.
False overused CRAM allocation can be optimized so that potentially more
aggressive parallel distribution can be feasible.

For the example in Figure~\ref{fig:mv-alloc}, 32x8=256 wordlines are required for
the accumulated results (\texttt{c.cram}), 8x1=8 wordlines for the operand \texttt{a},
1 wordline for the operand \texttt{b}, and implicit 32 wordlines
for the intermediate results. In total, 256+8+32+1=297 wordlines,
which exceeds 256 wordlines of each array. Next, we will explain
how our optimizations make them fit. These optimizations mostly
take advantage of the divisible nature of bit-serial arithmetic ---
each bit of results is independently accessible.

\paragraph{Adaptive Precision} {This technique can
save memory space of the computed results.
The minimum feasible precisions are adopted to override the
precision in the original program:
Multiplying an $a$-bit and a $b$-bit number
is at most $(a+b)$ bits; accumulating $k$ $a$-bit
numbers requires only $a$+$\lceil\log(\text{k})\rceil$ bits.
Specific to the example shown in Figure~\ref{fig:mv-impl},
though the results are accumulated on \texttt{i32},
only \texttt{i26} is required.
The input operands are both \texttt{i8}, so the result
of multiplication is within \texttt{i16}.
We now have \texttt{p=1024} \texttt{i16} accumulated, in total
$\log_2$1024+16=26 bits for each accumulation.
Therefore, we in total saved (32-26)x8+16=64 wordlines,
so now only 249 wordlines are required.}

\begin{figure}[t]
    \centering
    \includegraphics[width=\linewidth]{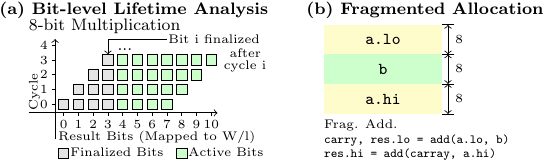}
    \caption{Bit-level lifetime, and fragmented allocation}
    \label{fig:frag-alloc}
\end{figure}

\paragraph{Lifetime Analysis}
The divisible nature of bit serial arithmetic makes
each bit have its own lifetime.
We extend the code scheduling in CHOPPER~\cite{chopper}
to an even broader applicability.
For a multiplication that is consumed immediately by an addition,
instead of keeping the whole 16 bits of multiplication, we can add it
to the accumulator as soon as a bit is finalized. As shown in
Figure~\ref{fig:frag-alloc}(a), after i cycles of multiplication,
the i-th bit is finalized, and
it always maintains a half-width
active window when doing a multiplication.
Therefore, this saves 16/2=8 wordlines,
and now we only occupy 243 wordlines.

\paragraph{Fragmented Allocation}
Fine-grain lifetime analysis will lead to fine grain memory recycling,
which makes fragmentation more likely.
This can hurt the utilization badly in conventional memory allocation.
However, the divisible nature of bit-serial PIM allows us to
allocate fragmented memory. 
We also need to divide operands and associated operations into
fragments, so that even tiny fragments can be utilized
when required.
Figure~\ref{fig:frag-alloc}(b) shows an example of fragmented allocation.
\texttt{a} is divided into two fragments straddled by \texttt{b},
and fragmented operations are generated for this allocation.

\paragraph{Data Loading \& Packing} Once
parallelism distribution passed the
constraint check after these optimizations,
the compiler will then inspect the memory access
pattern to generate code for data loading and packing.
If several tiles are using the same data, the common
memory traffic will be converted to on-chip network
communication. The compiler also analyzes the
the operands loaded to the CRAM arrays to determine
how they should be broadcast and shuffled through
the H-tree. For example, \texttt{xo} is not related to
reading \texttt{b.cram[y,k]}, so loaded \texttt{b}
should be broadcast to all the arrays mapped to different iterations of
\texttt{xo}.

\subsection{Implementation}

\rev{
We integrate our compiler analysis and transformations to the TVM compilation flow.
TVM provides rich code organization primitives to tune loop organizations and allocate memory buffers.
We start with the initial code organization provided by user, and apply our
explored parallelism and memory allocation. This is lowered to an IR with all
the loops and buffers instantiated. Then CRAM data optimization is done
on this level of IR. If the memory occupancy satisfies the hardware constraints,
all the arithmetic operations are rewritten in bit-serial intrinsics, including
arithmetic operations, memory loading, and data transfer, which are ready for code generation.
If not, we invoke a feedback loop to explore a more conservative code organization
for less memory occupancy.
}

\section{Evaluation Methodology} \label{sec:method}


\subsection{Modeling \sysname}

\paragraph{Performance}
We develop a cycle-accurate simulator 
in C++ to model the \sysname hardware. The simulator
executes a program written in the ISA described in Section \ref{section:isa}.
An input configuration file is used to specify various parameters of the microarchitecture.
Various metrics like cycles and energy breakdown by each component or instructions can also be reported by the simulator.

\paragraph{Area and energy model}
We develop an area and energy model to fairly compare \sysname
against our baseline, the NVIDIA A100.
We write Verilog RTL for the static H-tree network, shuffle logic, instruction controller, transpose unit, and register file.  For RAM blocks, we use the OpenRAM memory compiler \cite{open_ram}.
We verified and synthesized using Synopsys VCS and DC (using FreePDK45\cite{ncsu_freepdk45}) to obtain post-synthesis area and power. We further assume a 15\% area overhead for place and route
\cite{domain_specific_fpgas}.
For the PEs in each RAM, we write transistor level code and evaluate area and energy using SPICE with 22nm ASU PTM technology~\cite{asu_ptm}.
For the dynamic NoC, we use PAT-Noxim simulator \cite{pat_noxim} and extract area and energy values for the routers and links.
For the on-chip DRAM and PCIe controllers and transceivers, we obtain areas from A100 die analysis. 
For DRAM energy, we use a simple analytical model calibrated from memory-only microbenchmarks on the A100.
\rev{We scale all values, for both \sysname and A100, to 22nm using scaling factors for area, power and delay from \cite{Stillmaker201774}.}




\subsection{Configurations}
GPUs are the most common commercially available accelerators for DL workloads; so we compare \sysname against NVIDIA A100 GPU. Additionally, we compare against state-of-the-art prior
SRAM and DRAM based PIMs (DualityCache and SIMDRAM).
To make fair comparisons,
we build
three different \sysname configurations for each of the comparisons.

\paragraph{NVIDIA A100 GPU}
We provision \sysname to have the same area (825mm$^2$ in 7nm, i.e. 2950 mm$^2$ in 22nm) and DRAM bandwidth (12288 bits/cycle, i.e. 1866GB/s @1215 MHz).
GPU performance is measured by running on an A100 using NVIDIA's profiler NSight Compute. Each kernel is measured by averaging 500 launches to exclude the device overhead. To compare the dynamic energy of \sysname with A100, the static energy is normalized indirectly to A100 through having the same area footprint and DRAM bandwidth. 
Table \ref{tab:benchmarks2} describes the benchmarks we use and their
characteristics. 
5 highly parallel kernels from high performance libraries, including ArrayFire\cite{ArrayFire} (\texttt{fir}), and CUTLASS\cite{Kerr_CUTLASS_2022} (\texttt{gemm}, \texttt{gemv}, \texttt{conv2d}) are used as microbenchmarks. 
These microbenchmarks use integer datatypes. We choose different precisions for different benchmarks to demonstrate precision agnosticism.
We use quantized Resnet18 from MxNet Model Zoo to demonstrate the capability of targeting end-to-end workloads.

\begin{table}[t]
\centering
\caption{\rev{Benchmarks used for evaluation}}
\footnotesize
\setlength\tabcolsep{1.5pt}
\begin{tabular}{|l|p{3cm}|l|l||}
\hline
\rowcolor{LightBlue} \centering
\textbf{Benchmark} & \textbf{Size} & \textbf{Precision} & \textbf{Comparison} \\
\hline
vecadd & input=15728640 & int8 & A100\\
\hline
fir & input=7833600, \newline filter=32 & int16, acc=int16 & A100\\
\hline
gemv &  m=61440, k=2048, n=1 & int8, acc=int32& A100\\
\hline
gemm &  m=61440, n=32, k=2048 & int4, acc=int16& A100\\
\hline
conv2d & input=9x9x256x2, weights=3x3x256x256  & int8, acc=int32& A100\\
\hline
resnet18 &  input=224x224x3x1, \newline output=1000x1 & int8, acc=int32 & A100\\
\hline
backprop & input=65536x16 & fp32 & DC\\
\hline
dwt2d & input=1024x1024 & fp32 & DC\\
\hline
gausselim &  input=256x256 & fp32 & DC \\
\hline
hotspot & input=1024x1024 & fp32 & DC \\
\hline
hotspot3d & input=512x512 & fp32 & DC \\
\hline
vgg13 & input=224x224x3x1, \newline output=1000x1 & binarized & SIMDRAM\\
\hline
vgg16 & input=224x224x3x1, \newline output=1000x1 & binarized & SIMDRAM\\
\hline
lenet & input=32x32, \newline output=10x1 & binarized & SIMDRAM\\
\hline
\end{tabular}%
\label{tab:benchmarks2}
\end{table}

\paragraph{Duality Cache (DC)} 
DC has 1.14 million processing elements and runs at a frequency of 2.6 GHz.
We design a \sysname chip (\sysname-D) sized to match the compute throughput of DC for a fair comparison. 
\sysname-D has 30 tiles (organized in a 6x5 mesh).
The CRAM size for DC is the same as for \sysname (256 bitlines x 256 wordlines).
\rev{DC uses Rodinia \cite{che-rodinia} benchmarks and hence we use the same for this comparison.
}

\paragraph{SIMDRAM} We also compare \sysname to SIMDRAM \cite{simdram_asplos_2021}.
We use the 1 bank configuration mentioned in their paper.
We design a \sysname configuration (\sysname-S) with a lower number of processing elements in \sysname to match those in SIMDRAM. 
\sysname-S has 1 tile.
\rev{
We use 3 full Deep Neural Networks (LeNet, VGG-13, and VGG-16) as benchmarks for comparison since they were used in the SIMDRAM work.
}

\rev{
We do not build performance models of the DC and SIMDRAM architectures. 
Instead we obtain the raw runtimes for the benchmarks they used by directly reaching out to the authors, and used those for comparison.
}

\begin{figure}[t]
    \centering
    \includegraphics[width=0.9\linewidth]{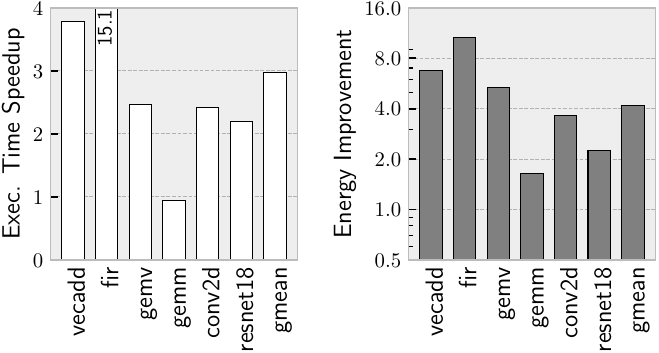}
    \caption{Comparing \sysname with NVIDIA A100}
    \label{fig:gpu-comparison}
\end{figure}

\begin{figure}[t]
\begin{subfigure}{0.49\linewidth}
\includegraphics[width=1\linewidth]{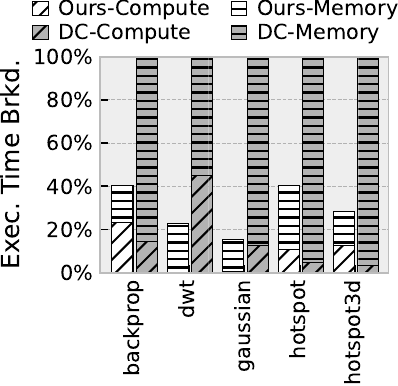}
\caption{\sysname-D vs. Duality Cache}
\end{subfigure}
\hfill
\begin{subfigure}{0.49\linewidth}
\centering
\includegraphics[width=0.75\linewidth]{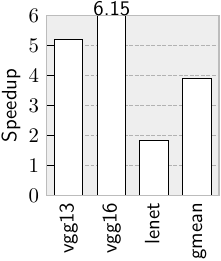}
\caption{\sysname-S vs. SIMDRAM}
\end{subfigure}
\caption{Appropriately provisioned \sysname compared with prior in-SRAM and in-DRAM systems.}
    \label{fig:simple_vs_simdram}
\end{figure}

\section{Results} \label{sec:eval}

\subsection{Comparison with state-of-art GPU}
Figure~\ref{fig:gpu-comparison} shows the execution time and energy
comparison against NVIDIA A100 GPU. On average, 
\sysname outperforms A100 by \gpuspeedup in execution
time, and \gpuenergy in energy.
The two main sources of the speedup are: (1) the high
fine-grain data parallelism in \sysname leads to reduced instruction overhead,
and the larger on-chip buffer
(512 MB on \sysname vs. 96 MB, including L2, shared memory and RF, on A100)
leads to more data reuse and reduces off-chip memory traffic.

\sysname significantly outperforms A100 on \texttt{fir} because of the unaligned
memory access caused by the sliding window.
In \sysname, this program behavior can easily be
handled by shifting bits across bitlines, while it prevents
the GPU from fully utilizing the memory bandwidth.
\sysname can achieve almost the same performance as A100 for \texttt{gemm}, even though A100 uses Tensor Cores for GEMM, which provide $2\times$ peak GOPs compared with \sysname.


\subsection{Comparison with SRAM PIM (Duality Cache)}
Fig~\ref{fig:simple_vs_simdram}(a) shows 
\sysname-D outperforms Duality Cache by \dcspeedup on average across several Rodinia benchmarks.
\sysname-D shows speedups over Duality Cache on \texttt{backprop}, \texttt{hotspot2d}, and \texttt{hotspot3d},
because of the tensor DSL programming compiler can easily analyze the memory footprint
and allocate buffers for memory reuse. In addition, Duality Cache still adopts a GPU-like
warp-wise execution, which imposes high overhead to coordinate unaligned data loading.
\sysname can simply shift across bitlines, even across CRAMs in a tile, so it outperforms DC on \texttt{dwt2d}.
\texttt{gaussian-elim} is bound by memory packing on DC, but our hardware is well specialized for it because of the H-tree, which also leads to fewer computational instructions.

\subsection{Comparison with DRAM PIM (SIMDRAM)}
Fig \ref{fig:simple_vs_simdram}(b) shows our comparison against SIMDRAM.
\sysname-S outperforms SIMDRAM~\cite{simdram_asplos_2021} by  \simdramspeedup on average across real world neural networks, because in-SRAM
processing takes advantage of data reuse in on-chip buffers. 
SIMDRAM has to pay DRAM read latencies for every computation and is at a disadvantage for workloads with data reuse.
\sysname's speedup is lower on LeNet
because the LeNet model is relatively small --- SRAM-DRAM transfer occupies a larger portion of execution, compared to the other networks.


\begin{figure}[t]
\begin{subfigure}{0.48\linewidth}
    \centering
    \includegraphics[width=0.95\linewidth]{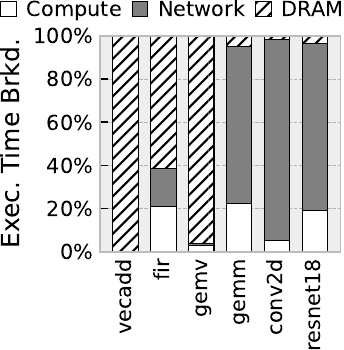}
    \caption{Execution time breakdown}
    \label{fig:time-brkd}
\end{subfigure}
\hfill
\begin{subfigure}{0.48\linewidth}
    \centering
    \includegraphics[width=0.95\linewidth]{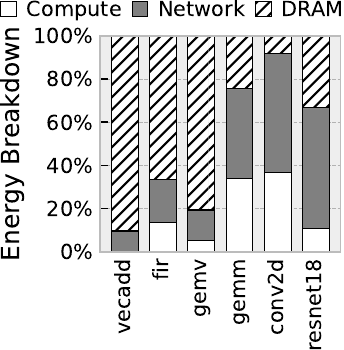}
    \caption{Energy breakdown}
    \label{fig:ener-brkd}
\end{subfigure}
\caption{Categorized breakdown of each workload.}\label{fig:breakdown}
\end{figure}

\subsection{Time and energy breakdown}
Figure \ref{fig:time-brkd} shows the breakdown of time spent in each benchmark. 
Since \texttt{vecadd} has low arithmetic intensity, most of the time is spent on DRAM loads and stores, as expected.
In \texttt{fir}, about 60\% of the time is spent on DRAM traffic.
\texttt{gemv} is also DRAM bound because of low reuse.
\texttt{gemm} and \texttt{conv2d} are dominated by network traffic, because
our compiler's loop organization transformation objective is to minimize
the estimated DRAM traffic by converting them to network data transfer.
\texttt{resnet18} is mainly a sequence of convolution
layers followed by elementwise operations.
We see that more time is spent on computation in \texttt{resnet18} than
a standalone convolution layer. This is because elementwise layers
(1) often have higher precision than convolution, and
(2) underutilize the hardware because of inter-CRAM
reduction.

Figure~\ref{fig:ener-brkd} shows the breakdown of energy consumed in each benchmark.
\texttt{vecadd}, \texttt{fir} and \texttt{gemv} are dominated by DRAM energy because of the limited reuse.
In microbenchmarks like \texttt{gemm} and \texttt{conv2d}, around 40\% of the energy is spent on computation.
For \texttt{resnet18}, even though 20\% of the time was spent on compute, 10\% energy is spent on computation. 

\begin{figure}[t]
    \centering
    \includegraphics[width=1\linewidth]{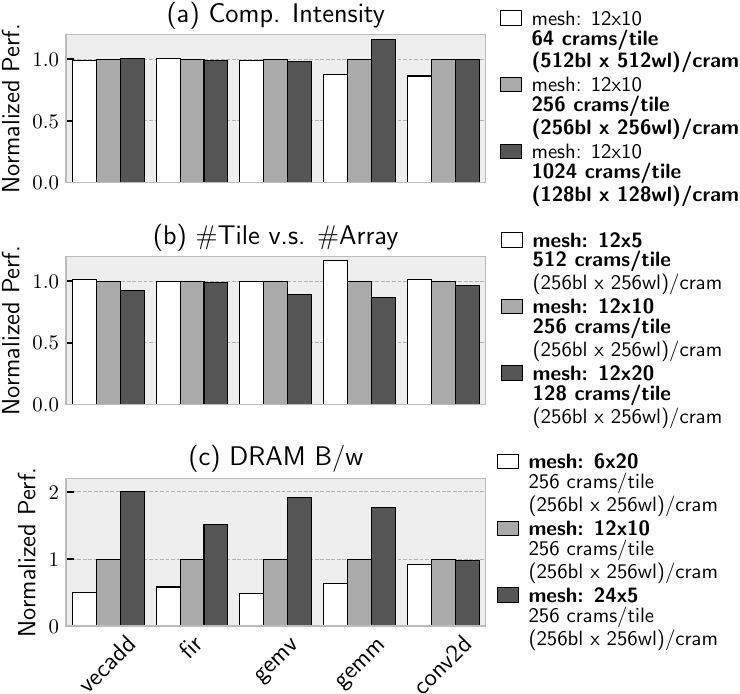}
    \caption{Performance sensitivity to hardware parameters}
    \label{fig:ubench-sens}
\end{figure}

\subsection{Sensitivity to different hardware parameters}
As shown in Figure~\ref{fig:ubench-sens}, we analyze a set of 7 different hardware configurations, obtained by varying 3 hardware parameters, with the microbenchmarks.
Figure~\ref{fig:ubench-sens}a studies the sensitivity
of number of compute resources (PEs) by tuning the size of each CRAM while retaining a constant
memory capacity.
Assuming each CRAM is a square (\#wordlines=\#bitlines), halving the number of bitlines
results in 4$\times$ more compute intensity (more PEs for the same amount of memory).
However, according to the cycle breakdown shown in Figure~\ref{fig:time-brkd},
the computations
occupy a small portion (on average less than 20\%)
of the execution time. So, increasing the compute resources without increasing memory capacity only slightly improves
the performance by 2.6\%, and decreasing
the compute resources slightly hurts performance by 5.4\%.

Figure~\ref{fig:ubench-sens}b studies the tradeoff
between the number of tiles and CRAMs per tile, while retaining 
the same number of compute resources. 
Increasing the number of tiles implies a larger dynamic network (NoC),
and increasing the number of CRAM per tile means more static network.
The results of this study suggest that
more tiles hurt performance by 8.2\%, and larger tiles provide dimishing returns (\textasciitilde1.5\% improvement).

Figure~\ref{fig:ubench-sens}c shows the tradeoff
by changing the memory bandwidth. This is achieved by changing the mesh geometry,
since only the top-row tiles have memory controllers.
The massive data parallelism requires
massive data to sustain. Therefore, workloads with poor reuse,
like \texttt{vecadd}, and \texttt{gemv}, which are bounded
by memory accesses, achieve nearly linear speedup
when doubling the memory bandwidth
(i.e number of columns in the mesh is doubled). 
Although according to Figure~\ref{fig:time-brkd}, \texttt{gemm}'s execution time
is not dominated by DRAM bandwidth, the performance is significantly improved when number of columns in the mesh increases.
This speedup is attributed to reduced data transfer time because the mesh height is reduced to half.
\texttt{conv2d} is an outlier; the performance is not affected (even slightly lowered)
by the memory bandwidth increase. Because of the shape of the kernel 
($3\times3\times256\times256$), we could only partition the memory load
across $3\times3=9$ tiles and broadcast them to all the tiles, which means
memory controllers are not fully utilized and loaded data is
broadcasted to further tiles because of the wider mesh width.
Overall, although more memory bandwidth is beneficial, for an iso-area and iso-memory-bandwidth comparison to the A100, we choose the configuration with $12\times10$ mesh as our main configuration.

\begin{figure}
    \centering
    \includegraphics[width=1\linewidth]{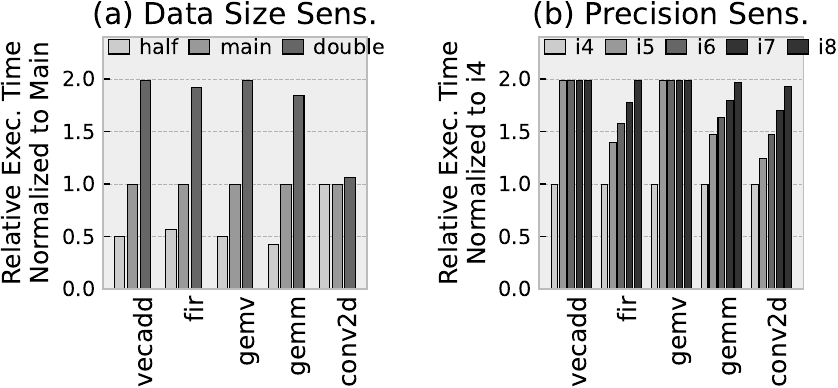}
    \caption{Perf. sensitivity to workload parameters}
    \label{fig:szpre-sens}
\end{figure}

\subsection{Sensitivity to different workload parameters}
Figure~\ref{fig:szpre-sens}(a) shows the sensitivity of \sysname's performance to workload sizes, by studying 2 additional sizes i.e.
halving and doubling the data.
The execution time of workloads with limited data reuse (e.g. \texttt{vecadd} and \texttt{gemv}) is linearly proportional to the data size. Same is the case with \texttt{gemm} and \texttt{fir}; they have better hardware utilization in larger sizes
because of data reuse.
Because of compute underutilization caused by shapes,
\texttt{conv2d} performance does not vary much with input size.

Figure~\ref{fig:szpre-sens}(b) shows the sensitivity of \sysname's performance to the precision of the inputs. 
A unique capability of bit-serial systems is to support any arbitrary precision. We vary the input precisions from 4-bits to 8-bits.  Since the DRAM representation always aligns to a power of 2, the DRAM traffic remains the same for int5 to int8.
Thus, the DRAM bound benchmarks (\texttt{vecadd} and \texttt{gemv}) exhibit the same performance under those precisions. 
Because computation and on-chip network traffic constitute a considerable amount of the execution time in \texttt{fir}, \texttt{conv2d}, and \texttt{gemm},
the performance of these workloads changes nearly linearly with precision.   
Note that adaptive precision eliminates the requirement to utilize 8-bit computations for smaller precisions.

\begin{figure}[t]
    \centering
    \includegraphics[width=0.7\linewidth]{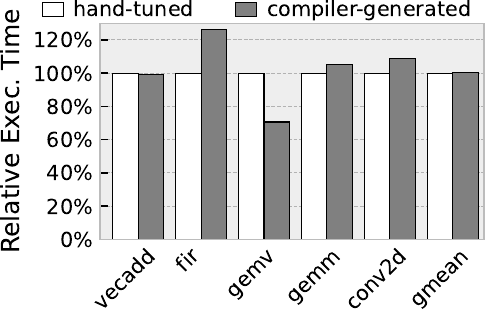}
    \caption{Comparing hand-tuned and compiler-generated 
    }
    \label{fig:manual_vs_compiled}
\end{figure}

\subsection{Quality of compiler-generated code}
With the goal of validating our compiler, Figure~\ref{fig:manual_vs_compiled} compares compiler-generated code
with hand-tuned code. The geomean performances are nearly the same.
\texttt{fir}, \texttt{gemm}, and \texttt{conv2d},
are moderately slowed down,
because our compiler conservatively generates
code for the synchronization between
receiving broadcasted data and computation,
which serializes these two phases.
Manually coded versions can fine tune this
part of overhead to reduce data transfer.
In \texttt{gemv}, our compiler outperforms the hand-tuned code,
because as discussed in Section \ref{sec:compiler},
the compiler avoids inter-tile communication for
compute reduction to save significant overhead
on NoC communication.
The hand-tuned code, however, uses a sub-optimal algorithm.

\subsection{Chip Area Distribution} \label{section:area_results}
\begin{figure}[t]
    \centering
    \includegraphics[width=0.95\linewidth]{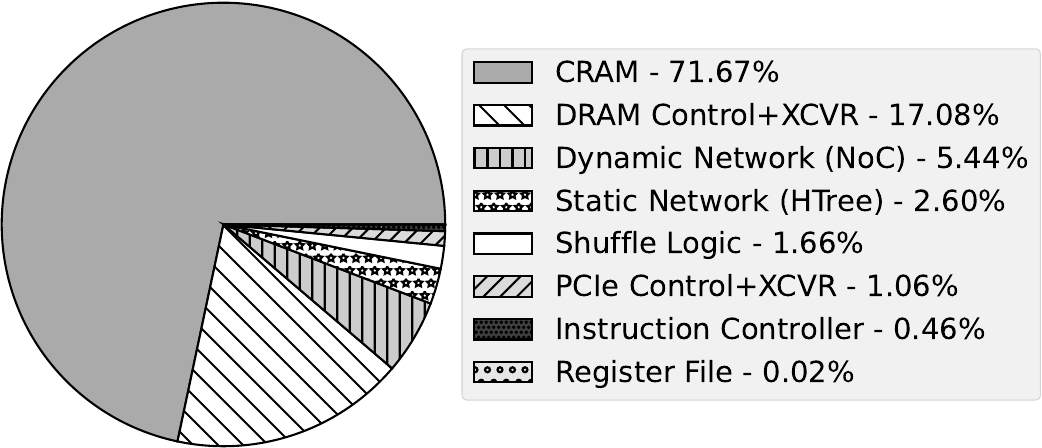}
    \caption{Chip area distribution of \sysname}
    \label{fig:area_division}
\end{figure}

Figure \ref{fig:area_division} shows the area distribution of the \sysname chip. 72\% of the chip area is consumed by the CRAMs, indicating a large percentage of useful compute/storage area.
The dynamic and static networks take $\sim$7.5\% of the chip area, while the shuffle logic occupies $\sim$1.5\% of the area.
The DRAM controller, transpose units and transceivers (XCVR) occupy 17\% of the chip. 
Considering the additional capabilities enabled by \sysname, the overhead is relatively low.

\section{Related Work} \label{sec:rel}

Instead of moving data to distant compute units,
PIM brings computation closer to the data. 
Recent works have used Non-Volatile Memories (NVM) like Resistive RAMs (ReRAM) or Spin-Transfer Torque Magnetic RAM 
(STT-MRAM)~\cite{isaac,prime,floatpim,wave_pim,cim_stt_mram,wave_pim}. NVM based solutions are nascent and are yet to reach large scale production,
and have endurance and technology scaling limitations.

Many DRAM-based PIM were proposed \cite{drisa,ambit,ComputeDRAM,simdram_asplos_2021},
without compilers, so they are difficult to program them.
CHOPPER~\cite{chopper} is a full-stack DRAM PIM which is programmed from a
bit-sliced DSL. Inspired by their code scheduling strategies,
we develop our bit-level lifetime analysis technique.
SRAM-based PIM has the advantage of simple integration with compute logic
using the same process, and also the ability to exploit data reuse in applications. 
\sysname uses SRAM-based PIM. Some SRAM-based approaches are analog \cite{compute_memory_uiuc,dima,conv_sram}, requiring expensive DACs and ADCs.
Other approaches use the property of enabling multiple wordlines in an SRAM at the same time \cite{neural_cache,duality_cache,compute_sram}. This requires reducing wordline voltage to avoid data corruption, and
modifying sense amplifiers. \sysname uses conventional dual ported RAMs instead, based on CoMeFa \cite{comefa}. This costs area, but is practical and robust.

In Neural Cache \cite{neural_cache} and Duality Cache \cite{duality_cache}, the focus is to repurpose existing caches in CPUs to perform in-situ computations. 
Neural cache uses an ad-hoc programming approach, and 
Duality Cache introduces a restricted version of the CUDA
programming interface; both of these are lower-level and expose 
hardware aspects to programmers (e.g. SRAM-array dimensions). 
Our Tensor DSL abstracts hardware and is easier to 
program and perform explorations with.

Other works such as PUMA \cite{PUMA_asplos_2019} and IMDPP \cite{imdpp_asplos_2018} develop compilers to make PIM systems easier to program. 
Their TensorFlow or C++ based graph-level programming interfaces are harder to perform performance tuning with than our Tensor DSL.
Also, they use ReRAM instead of SRAM-based PIM.

Recently, Processing-In-Memory has been proposed for FPGAs as well.
CCB~\cite{ccb} uses the same technology as Neural Cache~\cite{neural_cache} 
to enable block RAMs on an FPGA to perform computation, 
while CoMeFa~\cite{comefa} uses the dual-ported nature of block RAMs. 
Comparing with \sysname's Tensor DSL programming interface, these works still
require users to  design finite state machines to send
instructions to the RAM blocks, which is error-prone and time-consuming.

The bit-serial approach has a long history, going back to their use for neural networks in the 1980s~\cite{Murray_Smith_Butler_1987,Butler_Murray_Smith_1989}.
Stripes~\cite{stripes} is a more recent such DNN accelerator. \sysname combines a bit-serial approach with PIM.

\section{Conclusion} \label{sec:conc}

We present \sysname, a system for in-memory acceleration of massively parallel workloads like Deep Learning. 
Our system employs novel mechanisms for spatially-aware communication and bit-serial-aware computation.
While other PIM accelerators have been proposed for DL, our work makes significant strides in making PIM-based accelerators feasible for real-world DL problems. With the scalable hierarchical architecture combined with the H-tree and mesh interconnects at different levels, along with the shuffle network, adaptive precision and constant operation support, we make significant improvement in the capability of PIM-based accelerators.  
We demonstrate that \sysname can outperform state-of-the-art GPUs and PIM systems by 3$\times$ or more.

\bibliographystyle{IEEEtranS}
\bibliography{ref}

\end{document}